\documentclass[prd,twocolumn, nofootinbib, preprintnumbers]{revtex4}
\usepackage{graphicx, epsfig}
\usepackage{color}
\usepackage{mathrsfs}
\usepackage{amsmath}

\newcommand{\xv}{\vec{x}}
\newcommand{\grad}{\vec{\nabla}}
\newcommand{\kv}{\vec{k}}
\newcommand{\ra}{\rightarrow}

\begin{document}
\preprint{DESY 10-221}
\title{Gravity waves from the non-renormalizable Electroweak Vacua phase transition}
\author{Eric Greenwood}
\affiliation{CERCA, Department of Physics, Case Western Reserve University, Cleveland, OH 44106-7079}
\author{Pascal M.~Vaudrevange}
\affiliation{CERCA, Department of Physics, Case Western Reserve University, Cleveland, OH 44106-7079}
\affiliation{DESY, Notkestrasse 85, 22607 Hamburg, Germany}
 %%%%%%%%%%%%%%%%%%%%%%%%%%%%%%%%%%%%%%%%%%%%%%%%%%%%%%%
\begin{abstract}
  It is currently believed that the Standard Model is an effective low
  energy theory which in principle may contain higher dimensional
  non-renormalizable operators. These operators may modify the
  standard model Higgs potential in many ways, one of which being the
  appearance of a second vacuum. For a wide range of parameters, this
  new vacuum becomes the true vacuum. It is then assumed that our
  universe is currently sitting in the false vacuum. Thus the usual
  second-order electroweak phase transition at early times will be
  followed by a second, first-order phase transition. In cosmology, a
  first-order phase transition is associated with the production of
  gravity waves. In this paper we present an analysis of the
  production of gravitational waves during such a second electroweak
  phase transition. We find that, for one certain range of parameters,
  the stochastic background of gravitational waves generated by bubble
  nucleation and collision have an amplitude which is estimated to be
  of order $\Omega_{GW}h^2\sim10^{-11}$ at $f=3\times 10^{-4}$Hz,
  which is within reach of the planned sensitivity of LISA. For
  another range of parameters, we find that the amplitude is estimated
  to be of order $\Omega_{GW}h^2\sim10^{-25}$ around $f=10^3$Hz, which
  is within reach of LIGO. Hence, it is possible to detect gravity
  waves from such a phase transition at two different detectors, with
  completely different amplitude and frequency ranges.
\end{abstract}

\maketitle

\section{Introduction}

Recently, the authors in \cite{Greenwood:2008qp} have analyzed the
scalar field theory of a standard model Higgs field whose potential
includes non-renormalizable operators up to mass dimension $8$. In
this model, a second true vacuum appears which may be responsible for
the observed late-time acceleration of the universe. While the
electroweak phase transition remains second order, the subsequent
phase transition from the second false vacuum to the true vacuum is
first order, giving rise to the possibility of bubble nucleation. In
this picture, we are sitting in the second false vacuum (located e.g.
around $\phi\approx0.3$ TeV in Fig.~\ref{phi8_pot}), awaiting the
plunge into the true vacuum (around $\phi\approx0.8$ TeV). There are
certain fore-bearers of this impending doom that are in principle
observable.

In particular, it is well-known that bubble collisions lead to the
production of gravitational waves (GW), due to the breaking of
symmetry.  The strength of the phase transition, as well as the
amplitude of the GW are strongly model and parameter dependent. Thus
it seems worthwhile to investigate whether this non-renormalizable
Higgs model can have a background of interest for current and proposed
GW detectors\footnote{It should be pointed out that a possible
  detection of gravitational waves from bubble collisions of this
  phase transition would give us only a brief time before the joint
  bubble of true vacuum passes through us -- the bubble wall moves at
  close to and gravitational waves move exactly at the speed of
  light.}, specifically LIGO, Advanced LIGO, Geo 600, Virgo
\cite{Abadie:2010mt} and as well as the proposed LISA experiment
\cite{Cutler:1997ta}.

The two main sources of GW production are the nucleation and collision
of bubbles of true vacuum and the onset of turbulence. In this paper
we shall restrict ourselves to the first case, that of collisions of
bubbles of true vacuum. In order to tackle this problem, we rely on
the pioneering work of \cite{Kosowsky:1991ua, Kamionkowski:1993fg} who
developed a semi-analytical model for the GW spectrum emitted from
bubble collisions.

This paper is organized as follows. Section~\ref{sec:EM} briefly
reviews the model proposed in \cite{Greenwood:2008qp}. In
Section~\ref{sec:GW Prod}, we review the thin-wall approximation
of~\cite{Kosowsky:1991ua,Kamionkowski:1993fg}. In Section~\ref{sec:gwp}
we present the results of our calculation. Finally, we conclude in
Section~\ref{sec:conclusions}.

\section{Electroweak Model}
\label{sec:EM}

In \cite{Greenwood:2008qp} the authors considered the standard Higgs
potential with additional higher order non-renormalizable terms. These
non-renormalizable terms do not pose a problem as long as they are
suppressed by small enough coupling constants, i.e.~large enough
masses. The potential considered is given by,
\begin{equation}
  V(\phi)=-\frac{\mu^2}{2}\phi^2+\frac{\lambda_1}{4}\phi^4-\frac{\lambda_2}{8}\phi^6+\frac{\lambda_3}{16}\phi^8+V_0.
  \label{V1}
\end{equation}
For later use, we rewrite the potential as
\begin{equation}
  V(\phi)=-\varepsilon_0\phi_1^3\phi_2^3\phi^2+\frac{\lambda_3}{16}(\phi^2-\phi_1^2)^2(\phi^2-\phi_2^2)^2+V'_0\,,
  \label{V}
\end{equation}
with 
\begin{align}
  \phi_1^2&\equiv\frac{2}{\lambda_3\phi^2_2}\left(\lambda_1-\frac{1}{4}\frac{\lambda_2^2}{\lambda_3}\right),\\
  \phi_2^2&\equiv\frac{1}{2}\frac{\lambda_2}{\lambda_3}\left(1+\sqrt{3-\frac{8\lambda_1\lambda_3}{\lambda_2^2}}\right),\\
  \varepsilon_0&\equiv\frac{\mu^2}{2\phi_1^3\phi_2^2}-\frac{\lambda_3}{8\lambda_1\lambda_2}(\phi_1^2+\phi_2^2),\\
  V'_0&\equiv V_0-\frac{\lambda_3}{16}\phi_1^4\phi_2^4,
\end{align}
In the latter form, the finite temperature effective potential
formulation of the theory is more easily written down.

\begin{figure}[tb]
  \includegraphics{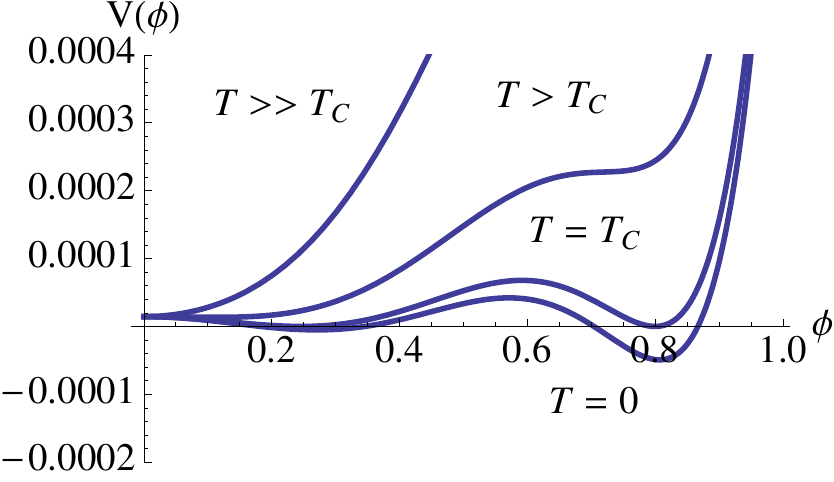}
  \caption{The temperature dependence of the $\phi^8$ Higgs field
    potential given in Eq.~(\ref{V_T}). The values used in this plot
    are $\phi_1=0.246$ TeV, $\phi_2=0.8$ TeV, $\lambda_3=0.154$
    TeV$^{-4}$, $V'_0=0$ and $\varepsilon_0=0.01$ TeV$^{-4}$.}
\label{phi8_pot}
\end{figure}

The parameter $\varepsilon_0$ assumes the role of a controllable fine
tuning of the potential. When $\varepsilon_0=0$, the potential in
Eq.~(\ref{V}) has four degenerate minima at
$\phi=\pm\phi_1,\pm\phi_2$. When $\varepsilon_0\not=0$, the degeneracy
of the vacua between $\phi=\phi_1,\phi_2$ ($\phi=-\phi_1,-\phi_2$) is
broken and there is an energy difference between the energy densities
of the two vacua, as shown in Fig.~\ref{phi8_pot} (at $T=0$). The
difference in energy is given as 
\begin{equation}
  \delta V=\varepsilon_0\phi_1^3\phi_2^3(\phi_2^2-\phi_1^2)\,,
\end{equation}
where $\phi_2>\phi_1$.

Finite-temperature effects are approximated by adding a thermal mass
term to the potential in Eq.~(\ref{V}), hence the potential takes the
form $V(\phi,T)=cT^2\phi^2/2+V(\phi,0)$, where $c$ is generated by the
quadratic terms that acquire a $\phi$-dependent mass in the
high-temperature expansion of the one-loop thermal potential. In
general there are also terms in the high-temperature thermal expansion
that are proportional to $T^4\phi^4$. However, it was argued in
\cite{Grojean:2004xa} that these terms only lead to small corrections to the
potential, therefore we shall ignore these contributions. In terms of
the temperature, the effective potential then becomes 
\begin{equation}
  V(\phi,T)=-\varepsilon(T)\phi_1^3\phi_2^3\phi^2+\frac{\lambda_3}{16}(\phi^2-\phi_1^2)^2(\phi^2-\phi_2^2)^2+V'_0,
  \label{V_T}
\end{equation}
where 
\begin{equation}
  \varepsilon(T)=\varepsilon_0-\frac{cT^2}{\phi_1^3\phi_2^3}.
  \label{e_T}
\end{equation}
Using the procedure outlined in 
\cite{Viswanathan:1978yy,2000csot.book.....V}, the 
constant $c$ is found to be given by
\begin{equation}
  c=\frac{1}{16}(3g^2+g'^2+4y_t^2+\frac{1}{32}\lambda_1)
  \label{c}
\end{equation}
where $g$ and $g'$ are the $SU(2)_L$ and $U(1)_Y$ gauge couplings, and
$y_t$ is the top Yukawa coupling. Using the excepted values of these
couplings and that the Higgs mass is $0.08$ TeV $<m_H<$ 0.15 TeV, one
finds that from Eq.~(\ref{c}), $c\approx1.7$. The critical temperature
is defined when $\varepsilon(T)=0$ is zero or in other words
\begin{equation}
  T_c^2=\frac{\varepsilon_0\phi_1^3\phi_2^3}{c}.
  \label{T_c}
\end{equation}
In this case, the vacua are degenerate, as discussed above.

In Fig.~\ref{phi8_pot} we plot the temperature dependence of the
potential in Eq.~(\ref{V_T}). Since the potential is symmetric for
$\phi\ra-\phi$, we will consider only the $\phi>0$ half-plane. For
very high temperatures ($T>>T_c$), the quadratic term of the potential
dominates and the potential consists of a characteristic ``U'' shape
with a single minimum at $\phi=0$. Here, the Higgs field has zero
expectation value and the electroweak symmetry is left unbroken. As
$T$ decreases, $\phi=0$ becomes a maximum ($T>T_c$) and the field
begins to roll down the potential toward the first minimum, which is
forming at $\phi=\phi_1$. The second minimum is forming at
$\phi=\phi_2$ at higher energy, making it non-accessible since it is a
less favorable state. The non-zero expectation value of the Higgs
field at $\phi=\phi_1$ breaks the electroweak symmetry. This is the
standard picture of electroweak symmetry breaking from a field going
through a second order phase transition. However, the Higgs potential
keeps evolving as the temperature decreases. The the second minimum at
$\phi=\phi_2$ drops to lower energies until it becomes the true global
minimum after the temperature falls below $T_c$, causing the minimum
at $\phi=\phi_1$ to be the false minimum. In other words, the Higgs
field sitting there in the false vacuum can undergo a first order
phase transition by tunneling to the global minimum at $\phi=\phi_2$
and forming bubbles of true vacuum.

As is well known, both nucleation of bubbles and their collisions can
source gravitational waves. In the next section we will review the
production of gravitational waves via bubble collision.

\section{Gravitational Wave Production in Phase Transitions}
\label{sec:GW Prod}

We are interested in production of gravitational waves from a
first-order phase transition of the Higgs field. The Lagrangian for
the Higgs field $\varphi$ is given by 
\begin{equation}
  L=\frac{1}{2}\partial^{\mu}\varphi\partial_{\mu}\varphi-V(\varphi)\,,
  \label{L}
\end{equation}
where the potential possesses at least two non-degenerate local
minima, such as that in Eq.~(\ref{V}) (as shown in
Fig.~\ref{phi8_pot}). In the following, we use a metric with signature
$(+---)$. Classically, the false vacuum state is stable: a field
sitting in this vacuum will remain there forever. However, quantum
effects can cause the false vacuum to decay to the true vacuum. This
decay proceeds via the nucleation and expansion of bubbles of true
vacuum, which spontaneously appear in the false vacuum. The bubbles
then expand due to the energy difference between the two vacua, which
induces an effective pressure on the bubble wall. This causes a bubble
to expand with an initial acceleration, with its expansion speed
quickly approaching the speed of light.

In \cite{Coleman:1977py}, Coleman showed that the bubble evolves
according to a Klein-Gordon equation with $O(3,1)$ symmetry which
implies that the position of the bubble wall is given by
\begin{equation}
  \xv_{wall}^2-t^2=R_0^2,
\end{equation}
where $R_0$ is the initial radius of the bubble and $\xv_{wall}$
denotes a fiducial point with in the bubble wall. When $t$ is
sufficiently large, we see that the position of the bubble wall is
given by $R(t)\approx t$, hence the radius of the bubble wall is
proportional to the elapsed time. From Eq.~(\ref{L}), the
stress-energy tensor associated with the bubble is
\begin{equation}
  T_{\mu\nu}(\xv,t)=\partial_{\mu}\varphi\partial_{\nu}\varphi-g_{\mu\nu}L
  \label{T}
\end{equation}
where the energy density of the bubble is the $T_{00}$ component. 

To determine the gravity-wave production we shall employ the envelope
approximation, originally developed by \cite{Kosowsky:1992vn} in the
context of the linearized gravity approximation. This approximation is
valid for bubble sizes less than $H^{-1}$. The energy radiated in
gravitational waves is given by the Fourier transform of the spatial
components of the stress-energy tensor in Eq.~(\ref{T}). Here, we will
neglect the $g_{\mu\nu}L$ term, since it is a pure trace term and does
not contribute to the production of the gravitational waves. Following
Weinberg \cite{Weinberg}, the Fourier transform can be shown to be 
\begin{equation}
  T_{ij}(\hat{\kv},\omega)=\frac{1}{2\pi}\int_0^{\infty}dt~e^{i\omega
    t}\int d\xv\partial_i\varphi\partial_j\varphi~e^{-i\omega\hat{\kv}\cdot\xv},
  \label{T_ij}
\end{equation}
where $\hat{\kv}$ is a unit wave vector. In the envelope
approximation, one assumes that the regions that have overlapped do
not contribute substantially to the gravitational radiation. Thus,
these regions are excluded from the spatial integrals. Assuming that
the time of nucleation for the bubbles is $t=0$,
\cite{Kosowsky:1992vn} showed that Eq.~(\ref{T_ij}) leads to
\begin{align}
  T_{ij}(\hat{\kv},\omega)=\frac{\rho_{vac}}{6\pi}\int_0^{\infty}dt~e^{i\omega t}&\sum_{n=1}^Nt^3e^{-i\omega\hat{\kv}\cdot\kv_n}\times\nonumber\\
    &\int_{S_n}d\Omega e^{-i\omega\hat{\kv}\cdot\kv}\hat{\kv}_i\hat{\kv}_j
\end{align}
where the sum is over the number of colliding bubbles. From Weinberg, the total energy radiated in gravity waves is given as
\begin{equation}
  \frac{dE}{d\omega d\Omega}=2G\omega^2\Lambda_{ij,lm}(\hat{\kv})T_{ij}^*(\kv,\omega)T_{lm}(\kv,\omega)
  \label{dE/dom}
\end{equation}
where $\Lambda_{ij,lm}$ is the projection tensor for gravity waves:
\begin{align}
  \Lambda_{ij,lm}(\hat{\kv})=&\delta_{il}\delta_{jm}-2\hat{\kv}_j\hat{\kv}_m\delta_{il}+\frac{1}{2}\hat{\kv}_i\hat{\kv}_j\hat{\kv}_l\hat{\kv}_m\nonumber\\
  &-\frac{1}{2}\delta_{ij}\delta_{lm}+\frac{1}{2}\delta_{ij}\hat{\kv}_l\hat{\kv}_m+\frac{1}{2}\delta_{lm}\hat{\kv}_i\hat{\kv}_j.
\end{align}

\begin{figure}[htp]
\includegraphics{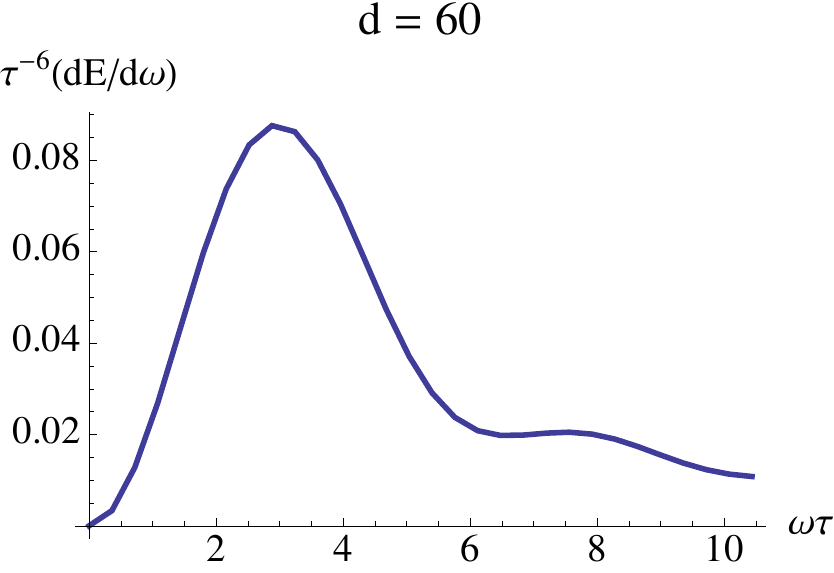}
\caption{Scaled energy spectrum for initial bubble separation of $d=60$.}
\label{ngwp}
\end{figure}

\begin{figure}[htp]
\includegraphics{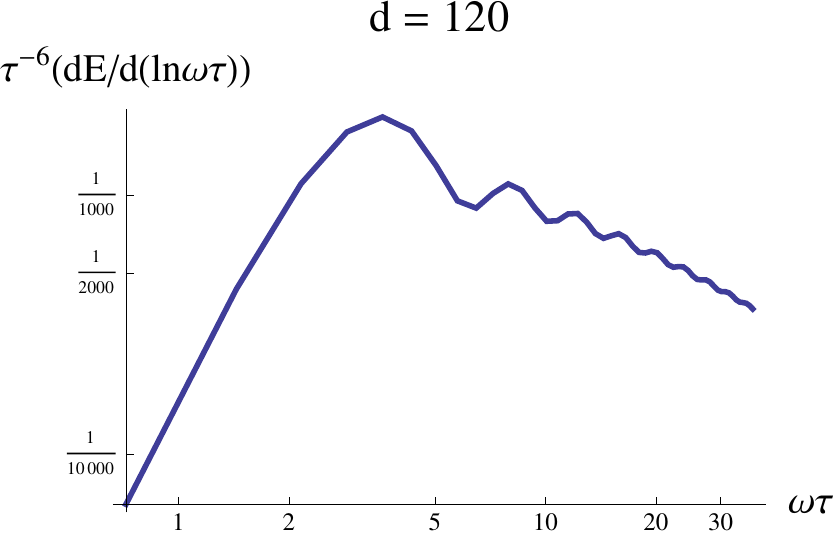}
\caption{Scaled energy spectrum per octave frequency interval.}
\label{ogwp}
\end{figure}

To compare our results with those found in \cite{Kosowsky:1992vn}, we
compute the case of two-bubbles colliding in the linearized gravity
approximation. In Figs.~\ref{ngwp} and \ref{ogwp} we plot some of our
numerical results. Here $d$ is the initial distance between the
two-bubbles. To simplify notation, we are plotting in units of
$G\rho_{vac}/3$ which is just a constant, thus the relative magnitude
of the curves in Figs.~\ref{ngwp} and \ref{ogwp} will be reduced. In
Fig.~\ref{ngwp} we plot the scaled energy spectrum for initial
separation of $d=60$, while in Fig.~\ref{ogwp} we plot the scaled
energy spectrum per octave frequency interval. Comparing our results,
we see that the results are qualitatively the same, hence consistent
with those in \cite{Kosowsky:1992vn}. This is expected since we are
using the envelope method, where the only dependence on the specific
models potential comes in the form of $\rho_{vac}$, which is just a
constant.

\section{Gravitational Wave Parameters}
\label{sec:gwp}
To estimate the spectrum of gravitational waves that are sourced by
this phase transition, we need to estimate the tunneling rate for the
decay of the false vacuum state to the true vacuum state as well as
the temperature scale.  We also need to go into some details of the
subsequent evolution of the growth of the bubbles. In particular, we
need an estimate of the velocity of the detonation front $\zeta$. 

\subsection{Tunneling Rate  and Temperature}
\label{sec:DPT}

The decay rate is suppressed by the exponential of the effective
action, $\Gamma=\Gamma_0e^{-S_E(t)}$. The time scale for the decay
$\beta$ given by
\begin{eqnarray}
  \beta&=&-\frac{dS_E}{dt}\Big{|}_{t=t_*}
\end{eqnarray}
where $t_*$ is the time when the phase transition occurs. This gives
\begin{eqnarray}
  S_E(t)&\approx&S_E(t_*)-\beta(t-t_*)\,, 
\end{eqnarray}
with the Euclidean action given by 
\begin{eqnarray}
  S_E&=&\int d\tau d\xv\left[\frac{1}{2}\left(\frac{d\varphi}{d\tau}\right)^2+\frac{1}{2}(\grad\varphi)^2+V(\varphi)\right]\,,
  \label{S}
\end{eqnarray}
where $\tau=it$ is the Euclidean (or Wick rotated) time coordinate.
The full, $O(4)$ symmetric solution of Eq.~(\ref{S}) and the
nucleation of bubbles was analyzed first by \cite{Coleman:1977py,
  Callan:1977pt} for a universe at zero temperature. At finite
temperature $T$, \cite{Linde:1977mm} pointed out that
the field theory should be taken periodic in $\tau$ with
period $T^{-1}$. Hence Eq.~(\ref{S}) must be written as
\begin{equation}
  S_E(t)=\int_0^{1/T} d\tau d\xv\left[\frac{1}{2}\left(\frac{d\varphi}{d\tau}\right)^2+\frac{1}{2}(\grad\varphi)^2+V(\varphi,T)\right]\,,
  \label{S_t}
\end{equation}
where $V(\varphi,T)$ is the temperature dependent effective potential.
At sufficiently large temperatures, i.e. when the integrand is
approximately time independent over time scales $T^{-1}$ , the
integration over $\tau$ is reduced simply to the multiplication of
$T^{-1}$, or $S_E=S_3/T$ \cite{Linde:1981zj}. Here $S_3$ has $O(3)$
symmetry and is given by
\begin{eqnarray}
  S_3&=&4\pi^2 \int dr \left[\frac{1}{2}(\partial_r \varphi)^2+V(\varphi,T)\right]\,\\
  &=&4\pi R(T)^2S_1(T)-\frac{4}{3}\delta V(T)\pi R(T)^3\,,
  \label{S3}
\end{eqnarray}
where the last equality assumes the thin-wall approximation.  $S_1(T)$
is the solution of Eq.~(\ref{S_t}) to zeroth order in $\delta
V(T)=\varepsilon(T)\varphi_1^3\varphi_2^3 (\varphi_2^2-\varphi_1^2)$
\begin{eqnarray}
  S_1(T)&=&\int_{\varphi_1}^{\varphi_2}d\varphi\sqrt{2V(\varphi,T)}\label{S_1}\\
  &=&\frac{1}{15}\sqrt{\frac{\lambda_3}{2}} (\phi_2-\phi_1)^3(\phi_1^2+3\phi_1\phi_2+\phi_2^2)\,,
\end{eqnarray}
which is independent of the temperature as all $T$ dependence is in
$\delta V$.

The temperature dependent bubble radius
\begin{eqnarray}
  R(T)&=&\frac{2S_1}{\delta V(T)}\,,
\end{eqnarray}
is obtained by minimizing $S_3$. Plugging this back in (\ref{S3}) we find
\begin{eqnarray}
  S_3&=&\frac{16\pi S_1^3}{3\delta V(T)^2}\,.
  \label{S3_tw}
\end{eqnarray}

Note that for this thin wall approximation to hold, the bubble radius
must be much larger than the thickness of the wall, $R\gg\left.
  (\partial_\varphi^2V(\varphi))^{-\frac{1}{2}}\right|_{\varphi=\varphi_2}$.
We can now estimate the time scale of the bubble nucleation process
$\beta$
\begin{eqnarray}
  \beta&=&\frac{16\pi^2}{3} S_1^3 \partial_T \frac{1}{\delta V(T)^2 T}\,\nonumber\\
  &=&\frac{16\pi^2}{3} S_1^3 \frac{4cT^2\varphi^2-\delta V(T)}{T^2\varphi^4\delta V(T)}\,.
\end{eqnarray}

The temperature $T_*$ at which the transition takes place is computed
in a two step process. 

First, we fix $\epsilon_0$ by requiring that the probability is much
smaller than unity for our Hubble patch to already have performed the
second phase transition during the lifetime of the universe or in
other words, by demanding that our Hubble patch is still in the false
vacuum at $\varphi_1$. Therefore we have
\begin{equation}
  t_H^4 H^2 e^{-S}=t^2e^{-S}\ll1\,
\end{equation}
which implies
\begin{equation}
  S\gg 2\ln\frac{t}{t_P}\approx 280\equiv S_{\text{crit}}\,,
\end{equation}
where $t_P$ is the Planck time. The smallest value of the action, i.e.
the highest probability for the transition from
$\phi_1\rightarrow\phi_2$ to occur, is realized at zero temperature.
Thus we can substitute the action at temperature $T$ by the zero
temperature action to obtain a lower limit on the action.
\begin{eqnarray}
  S_E(0)&=&\frac{27\pi^2}{2}\frac{S_1^4}{\delta V(0)^3}\\
  &=&\frac{27\pi^2}{2}\frac{S_1^4}{\epsilon_0^3 (\phi_1^3\phi_2^3 (\phi_2^2-\phi_1^2))^3}\,,
\end{eqnarray}
which should be larger than the critical action
\begin{equation}
  S_E(0)>S_{\text{crit}}\,.
\end{equation}
This implies
\begin{equation}
   \epsilon_0 <\frac{27\pi^2}{2 S_{\text{crit}}} \frac{S_1^4}{(\phi_1^3\phi_2^3 (\phi_2^2-\phi_1^2))^3}\,.
\end{equation}

Then, we estimate the transition temperature by equating
\begin{eqnarray}
  S_E&\approx&\frac{S_3}{T}\,,
\end{eqnarray}
and solving for $T$, which should at least give us a rough estimate.

Taking $\varphi_1=0.246$TeV, $\varphi_2=0.8$TeV,
$\lambda_3=0.154$TeV$^{-4}$ and $S_\text{crit}=425$, we find
$\varepsilon_0=0.15$TeV$^{-4}$ and a transition temperature of order
$T_*=0.0128$TeV.

From the transition temperature we can now determine $\beta/H_*$, the
quantity of interest (as we shall see in the next section). We first
need to make use of the fact that during radiation domination
\begin{eqnarray}
  H^2&=&\frac{1}{3}\rho=\frac{\pi^2}{90} N(T) T^4\,,\\
  dt&=&-\frac{1}{2\pi}\sqrt{\frac{45}{\pi N(T)}}\frac{dT}{T^3}\,,
\end{eqnarray}
where $N(T)$ is the effective number of particles, see e.g.
\cite{Linde:1993jz} for more details. Thus we have
\begin{equation}
  \frac{\beta}{H}=-\frac{1}{H}\frac{\partial S_E}{\partial t}=2\sqrt{2\pi}T\frac{\partial S_E}{\partial T}\approx 5T\frac{(\partial S_3/T)}{\partial T}\approx 1500\,,
\end{equation}
c.f. \cite{Apreda:2001us} who dropped the numerical prefactor.

\subsection{Growth of Bubbles}
\label{sec:Det}

In this subsection, we estimate several quantities pertaining to the
growth of the bubbles, notably the relativistic $\gamma$ of the bubble
wall.

An expanding bubble of true vacuum behaves very much like a detonation.
This was first examined in detail by \cite{Steinhardt:1981ct} and the 
to first order phase transitions by \cite{Kamionkowski:1993fg}.

In a detonation, high temperature gas enters the nucleated bubble with
supersonic velocity. This prevents a shock from preceding the bubble
wall. The velocity of the detonation front is given by
\begin{eqnarray}
  \xi_d&=&\frac{1/\sqrt{3}+\sqrt{\gamma^2+2\gamma/3}}{1+\gamma}
  \label{xi}
\end{eqnarray}
where the relativistic $\gamma$ is approximately the ratio of vacuum
energy to thermal energy. There are several ways to estimate its
value, but they all amount to numbers of the same magnitude.

The efficiency factor, or fraction of latent heat that goes into the
kinetic energy rather than thermal energy, is defined as
\begin{eqnarray}
  \kappa(\gamma)&=&\frac{1}{1+0.715\gamma}\left[0.715\gamma+\frac{4}{27}\sqrt{\frac{3\gamma}{2}}\right]\,.
  \label{kappa}
\end{eqnarray}

To determine the latent heat and vacuum energy associated with the
transition, we begin with the value of the potential at the broken
phase minimum which is also the difference in free energy density
between the two states of the system. We denote this by
\begin{eqnarray}
  B(T)&=&-V(v(T),T)\,,
\end{eqnarray}
where $v(T)$ is the value of the broken phase minimum. The vacuum
energy associated with the transition is then
\begin{eqnarray}
  {\cal{E}}_*&=&B(T)-TB'(T)\Big{|}_{T=T_*}.
  \label{vac_en}
\end{eqnarray}
This results in a relativistic factor $\gamma$ of
\begin{eqnarray}
  \gamma&=&\frac{30{\cal{E}}_*}{\pi^2g_*T_*^4}\sim11200\,.
\end{eqnarray}

Alternatively, we could estimate $\gamma$ just like in
\cite{Kosowsky:1992vn} to be
\begin{eqnarray}
  \gamma&=&\frac{\rho_{vac}}{T_*^4}\sim23650\,.
\end{eqnarray}
Both estimates for $\gamma$ resultin a very large number.  The wall
velocity then approaches speed of light, and the efficiency approaches
unity
\begin{equation}
  \xi_d\approx1\approx\kappa\,.
\end{equation} 
Hence we are justified in using the approach in Sec.~\ref{sec:GW
  Prod}.

One final remark is in order. In Sec.~\ref{sec:EM} we have arbitrarily
set $V'_0=0$. However, the chosen value of $V'_0$ effects the value of
the vacuum energy associated with the transition, see
Eq.~(\ref{vac_en}). In particular, the presence of a non-zero $V'_0$
acts as a overall shift in the vacuum energy, since the derivative of
the potential at the broken minimum, $B'(T)$, is independent of
$V'_0$. Therefore one would write the vacuum energy as
${\cal{E}}'_*={\cal{E}}_*+V'_0$.  However, since $\gamma$ is so large,
the constant shift, which is on the order of a few TeV, would increase
the above numbers. But as $\gamma, \xi_d,$, and $\kappa$ are already
indistinguishable from their values in the limit
$\gamma\rightarrow\infty$, this shift is not important for the
determination of the gravitational wave spectrum.

\subsection{Gravitational Wave Spectrum}
\label{sec:GWSpec}

The spectrum of gravitational waves emitted from bubble collisions was
first found by \cite{Kamionkowski:1993fg}. The fraction of the energy
density compared to the critical energy density is given by
\begin{align}
  \Omega_{GW}h^2\approx&1.1\times10^{-6}\kappa^2\left(\frac{H_*}{\beta}\right)^2\nonumber\\
  &\times\left(\frac{\gamma}{1+\gamma}\right)^2\left(\frac{\xi_d^3}{0.24+\xi_d^3}\right)\left(\frac{100}{g_*}\right)^{1/3}\,.\label{Oh2}
\end{align}
The maximum of the spectrum is located at
\begin{align}
  f_{max}\approx&5.2\times10^{-8}\text{Hz}\left(\frac{\beta}{H_*}\right)\left(\frac{T_*}{1\text{GeV}}\right)\left(\frac{g_*}{100}\right)^{1/6}\,,
\end{align}
and the characteristic amplitude of a gravity wave is approximately
\begin{align}
  h_c(f_{max})\approx&1.8\times10^{-14}\kappa\left(\frac{\gamma}{1+\gamma}\right)\left(\frac{H_*}{\beta}\right)^2\nonumber\\
  &\times\sqrt{\frac{\xi^3}{0.24+\xi^3}}\left(\frac{100}{g_*}\right)^{1/3}\label{hc}\,.
\end{align}
In the previous section we found that $\gamma\ra\infty$ so that
$\xi_d\ra1$ and $\kappa\ra1$. Plugging in these values, we find
\begin{align}
  \Omega_{GW}h^2\approx&1.1\times10^{-6}\left(\frac{H_*}{\beta}\right)^2\left(\frac{1}{1.24}\right)\left(\frac{100}{g_*}\right)^{1/3}\,,\\
  f_{max}\approx&5.2\times10^{-8}\text{Hz}\left(\frac{\beta}{H_*}\right)\left(\frac{T_*}{1\text{GeV}}\right)\left(\frac{g_*}{100}\right)^{1/6}\,,\label{f_max}\\
  h_c(f_{max})\approx&1.8\times10^{-14}\left(\frac{H_*}{\beta}\right)^2\sqrt{\frac{1}{1.24}}\left(\frac{100}{g_*}\right)^{1/3}\,.
\end{align}
Therefore for bubble collisions we get $\Omega
h^2\approx1.7\times10^{-11}$ and $h_c\approx3.1\times10^{-19}$,
peaking at a frequency around $f_{max}\approx3.5\times10^{-4}$ Hz.

We briefly note that for smaller values of
$\varepsilon_0\approx0.005$TeV$^{-4}$, the transition occurs at
temperature $T_*\approx3$GeV. This gives
$\beta/H_*\approx3.7\times10^9$, leading to an amplitude
$\Omega_{GW}h^2\sim1.1\times10^{-25}$ at peak frequency
$f_{max}\sim10^3$Hz.

\section{Conclusion}
\label{sec:conclusions}

We have considered the standard Higgs field, albeit with a potential
containing higher order non-renormalizable terms. Introducing these
non-renormalizable terms induced a second minimum in the potential,
see Fig.~\ref{phi8_pot}. The presence of the first quadratic term in
Eq.~(\ref{V_T}) ensures that the two minima are non-degenerate at
sufficiently low temperatures. There is a false-vacuum (a local
minimum) and a true-vacuum (a global minimum). In the context of this
model we assume that our universe is currently in the false vacuum.
The presence of the second lower minimum means that the false-vacuum
can decay via the nucleation and expansion of bubbles of true-vacuum
from a first-order phase transition. As the bubbles are growing, they
will eventually collide with other bubbles, resulting in the
production of gravity waves.  In this paper we have investigated the
gravity wave spectrum from bubble collisions using the envelope
approximation developed in \cite{Kamionkowski:1993fg}. We have also
considered the resulting gravity waves from detonating bubbles.

For bubble collision from a given first-order phase transition,
knowledge of the model-specific parameters $\beta$, $\xi$, $\kappa$
and $\gamma$ suffices to determine the resulting gravity wave
spectrum, see Eqs.~(\ref{Oh2}) - (\ref{hc}). The effective potential
for bubble nucleation suffices to determine the spectrum of
gravitational radiation. Here we find that for the Higgs potential
containing the non-renormalizable terms Eq.~(\ref{V}), the transition
occurs at a temperature $T_*\approx12.8$GeV (see Section
\ref{sec:DPT}), $\beta/H_*\approx300$, and $\gamma\sim10^4$, so that
$\xi_d\approx1\approx\kappa$ (see Section \ref{sec:Det}). Using these
results and Eqs.~(\ref{Oh2}) - (\ref{hc}) from Section
\ref{sec:GWSpec}, we find that the energy density of gravity waves
$\Omega h^2\approx1.7\times10^{-11}$, their characteristic amplitude
$h_c\approx3.1\times10^{-19}$, and a peak frequency around
$f_{max}\approx3.5\times10^{-4}$ Hz.

The question then arises whether or not the gravity wave signal is
potentially detectable by current or future gravity wave experiments.
Here we consider some of the possible detectors and their projected
sensitivities and compare with our results. First we shall consider a
current land-based detector. From
\cite{Thorne:1987af,Abramovici:1992ah} the ultimate sensitivity to a
background for the Laser Interferometric Gravitational Wave
Observatory (LIGO) is an amplitude around $2\times10^{-25}$ at a
frequency of 100Hz. However, as we saw in Section \ref{sec:GWSpec}, we
found that $f_{max}\approx8.2\times10^{-4}$Hz with an amplitude of
1.7$\times$10$^{-11}$, which are well below this requirement.
Therefore a detection by LIGO is very unlikely. Next we consider a
future space-based experiment. The peak sensitivity for the Laser
Interferometer Space Antenna (LISA)
\cite{Apreda:2001us,Delaunay:2006ws,Report} is reached in the
frequency range between $3\times 10^{-2}$Hz and $10^{-1}$Hz with
$\Omega_{GW}h^2\sim10^{-10}$ - $10^{-12}$. From Section
\ref{sec:GWSpec} we see that our results fall within this range of
detection from LISA, hence a detection by LISA could be possible given
our results.

One thing to note, however, is that the value of $\varepsilon_0=0.15$
TeV$^{-4}$ is an upper limit since smaller values of $\varepsilon_0$
further decrease the bubble nucleation rate. Repeating the
calculations above, we see from Eqs.~(\ref{Oh2}) - (\ref{hc}) that as
$\varepsilon_0$ decreases, $\Omega_{GW}h^2$ and $h_c(f_{max})$ both
decrease while $f_{max}$ increases. For the value of $\varepsilon_0$
given in Fig.~\ref{phi8_pot}, $\Omega_{GW}h^2\sim3\times10^{-18}$
which is well outside the range of LISA. However, as $\varepsilon_0$
falls below $0.005 \text{TeV}^{-4}$ the peak frequency and amplitude
shift into a window that may be detectable by LIGO. For this value of
$\varepsilon_0$ the transition occurs at temperature $T_*\approx3$GeV,
$\beta/H_*\approx3.7\times10^9$. This lead to an amplitude of
$\Omega_{GW}h^2\sim1.1\times10^{-25}$ at peak frequency
$f_{max}\sim10^3$Hz.

Even though we don't consider these here, additional contributions to
the gravity waves spectrum (which could be comparable to, or maybe
even stronger than bubble collisions) could come from the period of
turbulence after the phase transition, see \cite{Kosowsky:2001xp}.
First of all, turbulence itself sources gravitational waves. In
addition to that, turbulence can also source magnetic fields during
the time of the phase transition, see
\cite{Kosowsky:2001xp,Ahonen:1997wh}. These induced magnetic field
will then act as a source of gravitational radiation until they are
damped out. The gravitational radiation from the latter process is
subdominant to that of the turbulence, however, its peak frequency is
much higher and could potentially be detected at a land-based detector
such as LIGO. It would be interesting to investigate the spectrum of
gravity waves sourced by these effects as well as the induced magnetic
fields.

\section*{Acknowledgments}

The authors were supported by a NASA ATP grant as well as by a grant
from the US DOE to the theory group at CWRU. The research of PMV was
also supported by the ``Impuls und Vernetzungsfond'' of the Helmholtz
Association of German Research Centres under grant HZ-NG-603. We would
like to thank J.~T.~Giblin, Jr.  and D.~Stojkovic for useful
suggestions and discussions.

\bibliographystyle{unsrt}
\bibliography{GWS}

\end{document}